\documentclass[a4paper,12pt]{myarticle}
\usepackage{graphicx}
\usepackage{moreverb}
\renewcommand\figurename{\small {\sc Fig.}}
\renewcommand\tablename{\small {\sc Table}}
\begin{document}
\thispagestyle{empty}
\ \vspace{0cm}
\begin{center}
\renewcommand{\baselinestretch}{1.1} \normalsize
{\bf \large Noisy Covariance Matrices and Portfolio Optimization II} \par
\vspace{0.7cm}
{Szil\'ard Pafka$^{1}$ and Imre Kondor$^{1,2}$} \par \vspace{0.35cm}
{\it \small $^1$Department of Physics of Complex Systems, E\"otv\"os
University \\
P\'azm\'any P.\ s\'et\'any 1/a, H--1117 Budapest, Hungary} \par
\vspace{0.1cm}
{\it \small $^2$Market Risk Research Department, Raiffeisen Bank \\
Akad\'emia u.\ 6, H--1054 Budapest, Hungary} \par \vspace{0.45cm}
February, 2002 \par \vspace{1.1cm}
{\bf Abstract} \par \vspace{0.8cm}
\parbox{13cm}{\small
Recent studies inspired by results from random matrix theory
\cite{galluccio,bouchaud,stanley} found that covariance
matrices determined from empirical financial time series appear to
contain such a high amount of noise that their structure can
essentially be regarded as random. This seems, however, to be in
contradiction with the fundamental role played by covariance matrices in
finance, which constitute the pillars of modern investment theory and
have also gained industry-wide applications in risk management.
Our paper is an attempt to resolve this embarrassing paradox.
The key observation is that the effect of noise strongly depends on the
ratio $r = n/T$, where $n$ is the size of the portfolio and $T$ the
length of the available time series. On the basis of numerical
experiments and analytic results for some toy portfolio models we
show that for relatively large values of $r$ (e.g.\ 0.6) noise does,
indeed, have the pronounced effect suggested by
\cite{galluccio,bouchaud,stanley} and illustrated later by
\cite{bouchaud2,stanley2} in a portfolio optimization context,
while for smaller $r$ (around 0.2 or below), the error due to noise
drops to acceptable levels. Since the length
of available time series is for obvious reasons limited in any
practical application, any bound imposed on the noise-induced error
translates into a bound on the size of the portfolio.
In a related set of experiments we find that the effect of noise depends
also on whether the problem arises in asset allocation or in a risk
measurement context: if covariance matrices are used simply for
measuring the risk of portfolios with a fixed composition rather
than as inputs to optimization, the effect of noise on the measured risk
may become very small.
\par \vspace{0.3cm} {\it Keywords:} noisy covariance matrices,
random matrix theory, portfolio optimization, risk management
} \end{center} \vspace{1cm} \par
\rule{5cm}{0.4pt} \par
{\small {\it E-mail:} syl@complex.elte.hu (S.\ Pafka),
ikondor@raiffeisen.hu (I.\ Kondor)
\newpage \setcounter{page}{1}
\renewcommand{\baselinestretch}{1.1} \normalsize


\section{Introduction}

Covariance matrices of financial returns play a crucial role in several
branches of finance such as investment theory, capital allocation or risk
management. For example, these matrices are the key input parameters to
Markowitz's classical portfolio optimization problem \cite{markowitz},
which aims at providing a recipe for the composition of a portfolio of
assets
such that risk (quantified by the standard
deviation of the portfolio's return) is minimized for a given level of
expected return.
For any practical use of the theory it would therefore be necessary to
have reliable estimates for the volatilities and correlations of the
returns on the assets
making up the portfolio (i.e.\ for the elements of the covariance matrix),
which are usually obtained from historical return series.
However, the finite length $T$ of the empirical time series
inevitably leads to the appearance of noise (measurement error) in the
covariance matrix estimates.
It is clear that this noise becomes
stronger and stronger with increasing portfolio size $n$, until at a
certain $n$ one overexploits the available information to such
a degree that the positive definiteness of the covariance matrix (and
with that the meaning of the whole exercise) is lost.

This long known difficulty has been put into a new light by
\cite{galluccio,bouchaud,stanley} where the problem has been
approached from the point of view of random matrix theory. These
studies have shown that empirical correlation
matrices deduced from financial return series contain such a high amount of
noise that, apart from a few large eigenvalues and the corresponding
eigenvectors, their structure can essentially be regarded as random.
In \cite{bouchaud}, e.g., it is reported that about 94\% of the
spectrum of correlation matrices determined from return series on
the S\&P 500 stocks can be fitted by that of a random matrix. One wonders how,
under such circumstances, covariance matrices can be of any use in finance.
Indeed, in \cite{bouchaud} the authors
conclude that ``Markowitz's portfolio optimization
scheme based on a purely historical determination of the correlation matrix
is inadequate''.

Two subsequent studies \cite{bouchaud2,stanley2} found that
the risk level of optimized portfolios could be improved if
prior to optimization one filtered out the lower part of the eigenvalue
spectrum of the covariance matrix, thereby removing the noise
(at least partially). In both of these studies,
portfolios have been optimized by using the covariance matrix
extracted from the first half of the available empirical sample,
while risk was measured as the standard deviation of the return on these
portfolios in the second half of the sample.
\cite{bouchaud2,stanley2}
found a significant discrepancy between ``predicted`` risk
(as given by the standard deviation of the optimal portfolio in the
first half of the sample) and ``realized`` risk (given by its
actual realization in the second half),
although this discrepancy could be diminished
by the use of the filtering technique.
While these results suggest potential applications of
random matrix theory, they also reinforce the
doubts about the usefulness of empirical covariance matrices.

On the other hand, Markowitz's theory is one of the pillars of
present day finance. For example, the Capital Asset Pricing Model
(CAPM), which plays a kind of benchmark role in portfolio management,
was inspired by Markowitz's approach; various techniques of capital
allocation are based on similar ideas. Furthermore, over the years,
covariance matrices have found industry-wide applications also in risk
management. For example, RiskMetrics \cite{riskmetrics},
which is perhaps the most widely accepted methodology
for measuring market risk, uses covariance matrices
as its fundamental inputs.
The presence of such a high degree of noise in empirical covariance
matrices
as suggested by \cite{galluccio,bouchaud,stanley,bouchaud2,stanley2}
and the fact that these matrices are so widely
utilized in the financial industry constitute an intriguing paradox.

The motivation for our previous study \cite{pafkakondor} stemmed from this
context. In addition to the noise due to the finite length of time series,
real data always contain additional sources of error (non-stationarity,
changes in the composition of the portfolio, in regulation, in fundamental
market conditions, etc.). In order to get rid of these parasitic effects,
we based our analysis on data artificially generated from some toy models.
This procedure offers a major advantage in that the ``true`` parameters
of the underlying stochastic process, hence also the statistics of
the covariance matrix are exactly known. Furthermore, with a
comparison to empirical data in mind, where the determination of expected
returns becomes an additional source of uncertainty, we confined ourselves
to the study of the minimal risk portfolio. Our main finding was that
for parameter values typically encountered in practice the ``true`` risk
of the minimum-risk portfolio determined in the presence of noise
(i.e.\ based on the
covariance matrix deduced from finite time series) is
usually no more that 10--15\% higher than that of the portfolio
determined from the ``true`` covariance matrix.

In the present work we continue and extend our previous analysis, but
keep to the same toy-model-based approach as before. These models can
be treated both numerically and, in the limit when $n$ and $T$ go to
infinity with $r = n/T = \textrm{fixed}$, analytically. Varying the
ratio $r = n/T$ we show that the difference between ``predicted``
and ``realized`` risk can, indeed, reach the high values found in
\cite{bouchaud2,stanley2}
when $r$ is chosen as large as in those papers, but decreases
significantly for smaller values of this ratio. This observation eliminates
the apparent contradiction between \cite{bouchaud2,stanley2} and
our earlier results \cite{pafkakondor}.
Since in the simulation framework we know the exact process, not only
its finite realizations, we can compare the ``predicted`` and ``realized``
risk to the ``true`` risk of the portfolio. We find that ``realized``
risk is a good proxy for ``true`` risk in all cases of practical
importance and that ``predicted`` risk is always below, whereas
``realized``
risk is above the ``true`` risk. For asymptotically small values of
$n/T$ all the noise vanishes, but the value of $T$ is, for evident reasons,
limited in any practical application, therefore any bound one would
like to impose on the effect of noise translates, in fact, into a
constraint on the portfolio size $n$.

Regarding one other aspect of the problem, we find that the effect of
noise is very different depending on whether we wish to optimize the
portfolio, or merely want to measure the risk of a given, fixed portfolio.
While in the former case the effect of noise remains important up to
relatively small values of $n/T$, in the latter case it becomes
insignificant much sooner. This explains why covariance matrices
could have remained a fundamental risk management tool even to date.


\section{Results and Discussion}

We consider the following simplified version of the classical portfolio
optimization
problem: the portfolio variance $\sum_{i,j=1}^n w_i\,\sigma_{ij}\,w_j$
is to be minimized under the budget constraint $\sum_{i=1}^n w_i=1$, where
$w_i$ denotes the weight of asset $i$ in the portfolio and $\sigma_{ij}$
represents the covariance matrix of returns. One could, of course, impose
additional constraints (e.g.\ the usual one on the return), but this
simplified form provides the most convenient laboratory to test the effect
of noise,
since it eliminates the uncertainty coming from the determination of
expected returns. The solution to the optimization problem can then be
found
using the method of Lagrange multipliers, and after some trivial algebra
one
obtains for the weights of the optimal portfolio:
\begin{equation}
\label{eq:sol}
w_i^*=\frac{\sum_{j=1}^n \sigma_{ij}^{-1}}{\sum_{j,k=1}^n \sigma_{jk}
^{-1}}.
\end{equation}

Starting with a given ``noiseless`` covariance matrix $\sigma_{ij}^{(0)}$
we
generate ``noisy`` covariance matrices $\sigma_{ij}^{(1)}$ as
\begin{equation}
\sigma_{ij}^{(1)}=\frac{1}{T} \sum_{t=1}^{T} y_{it}\,y_{jt},
\end{equation}
where $y_{it}=\sum_{j=1}^n L_{ij}\,x_{jt}$, with
$x_{jt}\sim \textrm{i.i.d.\ N}(0,1)$ and $L_{ij}$ the
Cholesky decomposition of the matrix $\sigma_{ij}^{(0)}$
(a lower triangular matrix which satisfies
$\sum_{k=1}^n L_{ik}\,L_{jk}=\sigma_{ij}^{(0)}$, or
$L\,L^T=\sigma^{(0)}$).
In this way we obtain ``return series`` $y_{it}$ that have a
distribution characterized by the ``true`` covariance matrix $\sigma_{ij}
^{(0)}$, while
$\sigma_{ij}^{(1)}$ will correspond to the ``empirical`` covariance matrix.
Of course, in the limit $T\to\infty$ the
noise disappears and $\sigma_{ij}^{(1)}\to\sigma_{ij}^{(0)}$.
The main advantage of this simulation approach over empirical studies is
that the ``true`` covariance matrix is exactly known.

For our experiments we choose two simple forms for
$\sigma_{ij}^{(0)}$. First, we perform our simulations with the simplest
possible form for $\sigma_{ij}^{(0)}$, the identity matrix (Model I). In
order to move a little closer to the observed structures, however, we also
perform experiments with matrices $\sigma_{ij}^{(0)}$ which have one
eigenvalue chosen to be about 25 times larger than the rest and with the
corresponding eigenvector (representing the ``whole market``) in the
direction of $(1,1,\ldots,1)$, while keeping the simplicity of the identity
matrix in the other directions (Model II). This latter is meant to be a
caricature of the
covariance matrices deduced from financial return series (see
\cite{bouchaud,stanley}).

In order to see the effect of noise on the portfolio optimization
problem we compare the square roots of the following quantities:
\begin{enumerate}
\item $\sum_{i,j=1}^n w_i^{(0)*}\,\sigma_{ij}^{(0)}\,w_j^{(0)*}$,
the ``true`` risk of the optimal portfolio without noise, where
$w_i^{(0)*}$ denotes the solution to the optimization problem
without noise;
\item $\sum_{i,j=1}^n w_i^{(1)*}\,\sigma_{ij}^{(0)}\,w_j^{(1)*}$,
the ``true`` risk of the optimal portfolio determined in the
case of noise, where $w_i^{(1)*}$ denotes the solution to the
optimization problem in the presence of noise;
\item $\sum_{i,j=1}^n w_i^{(1)*}\,\sigma_{ij}^{(1)}\,w_j^{(1)*}$,
the ``predicted`` risk (cf.\ \cite{bouchaud2,stanley2}), that is
the risk that can be observed if the optimization is based on
a return series of length $T$;
\item $\sum_{i,j=1}^n w_i^{(1)*}\,\sigma_{ij}^{(2)}\,w_j^{(1)*}$,
the ``realized`` risk (cf.\ \cite{bouchaud2,stanley2}), that is
the risk that would be observed if the portfolio were held one more
period of length $T$, where $\sigma_{ij}^{(2)}$ is the covariance
matrix calculated from the returns in the second period.
\end{enumerate}

To facilitate comparison, we calculate the ratios of the square
roots of the three latter quantities to the first one, and
denote these ratios by $q_0,q_1$ and $q_2$, respectively.
That is $q_0,q_1$ and $q_2$ represent the ``true, ``the``predicted`` resp.\
the
``realized`` risk, expressed in units of the ``true`` risk in the
absence of noise.
For both model covariance matrices $\sigma_{ij}^{(0)}$,
we perform simulations for different values of the number of assets
$n$ and length of the time series $T$. For each given $n$ and $T$, we
generate several return series and covariance
matrices, and each time we calculate the corresponding ratios
$q_0,q_1$ and $q_2$. Finally, we calculate the mean and standard
deviation of these quantities for given $n$ and $T$.

\vspace{-0.1cm}
\begin{table}[!h]
\caption{``True``, ``predicted``, and ``realized`` risk for different
values of the number of assets $n$ and length of time series $T$
(the figures in parentheses denote standard deviations).
\vspace{0.2cm}
\label{tbl:simul}}
\begin{tabular}{*{9}{|c}|} \hline
Mod. & $n$ & $T$ & $q_0$ & $q_1$ & $q_2$ & $q_2/q_0$ & $q_2/q_1$ \\ \hline
\hline
I & 100 & 600  & 1.09 (0.01) & 0.92 (0.03) & 1.09 (0.03) & 1.00 (0.03) &
1.19 (0.05) \\ \hline
I & 500 & 3000 & 1.09 (0.01) & 0.91 (0.01) & 1.10 (0.02) & 1.00 (0.01) &
1.20 (0.03) \\ \hline \hline
I & 500 & 1500 & 1.22 (0.01) & 0.81 (0.01) & 1.22 (0.02) & 1.00 (0.01) &
1.49 (0.03) \\ \hline
I & 500 & 750  & 1.73 (0.07) & 0.57 (0.02) & 1.74 (0.07) & 1.00 (0.02) &
3.00 (0.16) \\ \hline \hline
II & 500 & 3000 & 1.09 (0.01) & 0.91 (0.01) & 1.10 (0.01) & 1.00 (0.01) &
1.20 (0.02) \\ \hline
II & 500 & 750  & 1.72 (0.06) & 0.58 (0.02) & 1.72 (0.08) & 1.00 (0.02) &
2.97 (0.17) \\ \hline
\end{tabular}
\end{table}
\vspace{-0.2cm}

The results of our simulations are given in Table \ref{tbl:simul}.
It can be seen that $q_0$ is higher than 1 for all values of $n$ and $T$,
as one would expect since the ``optimal`` portfolio obtained from
the ``noisy`` covariance matrix must be less efficient than the one
obtained from the ``true`` covariance matrix. It is also clear that $q_2$
is always very close to $q_0$, which suggests that ``realized`` risk
can be used as a good proxy for the ``true`` risk when the ``true``
covariance matrix is not known. This is the very case in empirical
studies, and therefore on the basis of
our simulation results we can expect that the values obtained for
``realized`` risk for example in \cite{bouchaud2,stanley2} must
indeed be close to the true risk figures.

This, unfortunately, fails to be true for the predicted risk.
As seen from the table, $q_1$ is always smaller than $q_0$ (or $q_2$).
Since $q_1$ (actually the
numerator of the fraction that determines $q_1$) is the only
risk figure that can be obtained in a framework when one sets up the
optimal portfolio based on a finite sample of returns (i.e.\ when the
true covariance matrix is unknown) and $q_0$ is the ``true`` risk of the
obtained portfolio, in such cases one will end up with underestimating
risk.
Therefore, optimization in the presence of noise will bias risk measurement
and lead to the underestimation of the risk of the optimal portfolio.
These conclusions are in perfect qualitative agreement with those in
\cite{bouchaud2,stanley2}.
The important point is the magnitude of the effect, however.
It is obvious, and also born out by Table \ref{tbl:simul}, that the
effect of noise should decrease with $r = n/T$, i.e.\ the risk measures
$q_0$ and $q_1$ should converge to 1 as the length of the time series
goes to infinity, with the size of the portfolio kept constant.

For $T,n\to\infty$ and $r=n/T=\textrm{fixed}$ one can, in fact, calculate
$q_0$ and $q_1$ analytically. Since the variance of a portfolio is a
rotation invariant scalar, it can be evaluated in the principal axis
system of the covariance matrix.
In terms of the eigenvalue density
\begin{equation}
\rho(\lambda)=\frac{1}{2\pi r}\frac{\sqrt{(\lambda_{+}-\lambda)
(\lambda-\lambda_{-})}}{\lambda}
\end{equation}
of the covariance matrix (see \cite{crisanti}), where
$\lambda_{\pm}={(1\pm\sqrt{r})}^2$, $q_0$ can be written
\cite{gabor} as
\begin{equation}
\label{eq:q0anal}
q_0=\frac{\sqrt{\int \rho(\lambda)/\lambda^2\;\textrm{d}\lambda}}{\int
\rho(\lambda)/\lambda\;\textrm{d}\lambda}.
\end{equation}
Simple integration yields $q_0=1/\sqrt{1-r}$. Similarly, for
$q_1$ one can obtain $q_1=\sqrt{1-r}$. It is easy to verify that
these asymptotic formulae fit the simulation results very well.
Also note that, as we have argued in \cite{pafkakondor}, when we have
sufficient information about the portfolio, i.e.\ when $n/T$ is small
enough (such as for
example in the first two rows in the table, or less), then $q_0$ is not
dramatically higher that 1, i.e.\ the inefficiency introduced by noise may
not necessarily be very large.
We have repeated our numerical experiments for Model II and obtained very
similar results (see the last two rows of Table~\ref{tbl:simul}).

In our view, the main message to be inferred from the above analysis is the
bound it implies for the noise-induced error. According to the above
formulae for $q_0$ and $q_1$, within the framework of our toy Model I the
error in the risk estimate is about $r/2$ for small $r$ (and we know that
predicted risk is always smaller than the true risk!) As a result of
inevitable additional imperfections (non-stationarity, deviation from normal
statistics, etc.), the error in the risk estimate of real-life portfolios
can only be larger than this, so $r/2$ can be viewed as a lower bound on
the error. Conversely, if we set a value for the acceptable error, we have
a bound on the ratio $n/T$. Since the length of meaningful time series is
always limited (by changes in the composition of the portfolio, changes in
the regulatory environment, in market conditions, etc.), this means that we
have an upper bound on the size of the portfolio whose risk can be
estimated with the pre-determined error. The filtering technique proposed 
in \cite{bouchaud2,stanley2} can then be regarded as a tool to break
through this upper bound.

Next, we compare our simulation results with results obtained for
covariance matrices determined from empirical return series. To
accomplish this, we use daily return series on 400 major US stocks
during the period
1991--1996 (1200 observations for each stock).
The data have been extracted from the the same dataset as
in \cite{bouchaud2}.\footnote{
We thank J.-P.\ Bouchaud and L.\ Laloux for making the dataset available
for us.}
Four non-overlapping samples of $n=100$ stocks
were created and divided into two periods of length $T=600$.
For each sample, the first period was used for optimization and
the calculation of ``predicted`` risk, while the second period for
determining ``realized`` risk. We found that the ratio of ``predicted``
and ``realized`` risk $q_2/q_1$ was $1.36\pm 0.07$, somewhat higher than
in our simulations for the same $n$ and $T$ (1.19).
To see whether this was caused simply by an
increase in volatility in the second period or not, we repeated our
calculations with swapping the two periods in each sample, but we
obtained similar results ($1.44\pm 0.08$). The additional bias in
$q_2/q_1$ could be caused, however, by volatility dynamics (e.g.\
conditional heteroskedasticity). In order to diminish the effect
of possible inhomogeneities in volatility we divided the sample such that
every other value from the return series was allocated to the first
``period`` while the rest to the second. In this case we obtained
$q_2/q_1=1.22\pm 0.06$, a result in line with our simulations.
In view of these latter findings, the results obtained e.g.\ in
\cite{bouchaud2} can be easily understood. In the empirical example of
\cite{bouchaud2} with $n=406$ and $T=654$, ``realized`` risk exceeded
``predicted`` risk by a factor of around 3, which seems reasonable if one
takes into account that
for this case the simulations, or the theory of the toy model, 
predict a value of 2.6.

Finally, we studied the effect of noise on portfolios which
were selected somehow independently from the covariance matrix data.
For example, one could invest in equal proportions in each asset,
or concentrate the portfolio to a few assets (e.g.\ stocks from
one single sector). We analyze therefore the implications of noise on
the risk measurement of portfolios whose weights are
determined independently of the covariance matrix data.
Let us consider a portfolio with weights $w_i$ fixed ($\sum_{i=1}^n w_i=1
$).
In a simulation setup in which covariance
matrices are generated as before, we compare
the ``true`` risk of the portfolio (as measured by the square root of
$\sum_{i,j=1}^n w_i\,\sigma_{ij}^{(0)}\,w_j$) to the ``observed`` risk
(deduced from $\sum_{i,j=1}^n w_i\,\sigma_{ij}^{(1)}\,w_j$), and
let $q$ denote the ratio of the second to the first.
Our simulations show that for large enough time series length $T$,
this quantity is very close to 1, no matter whether the number of
assets $n$ is small or large. For example, in the case in which the
``true`` covariance matrix is the $n\times n$ identity matrix and
the weights $w_i$ are chosen to be $1/n$ for all $i$, for
$T=100$ and $n=10$, one obtains $q=1.00\pm 0.07$, for $T=1000$ and
$n=10$, $q=1.00 \pm 0.02$, while for $T=100$ and $n=100$, $q=1.00
\pm 0.07$ (the mean is 1 and the standard deviation
is of the order of $1/\sqrt{T}$). The results obtained for other choices of
the
covariance matrix $\sigma_{ij}^{(0)}$ and of the weights $w_i$ are similar,
therefore we are led to the conclusion that the effect of randomness on the
risk estimate of portfolios with fixed weights is very limited.


\newpage
\section{Conclusion}

In this paper we have studied the implications of noisy covariance
matrices on portfolio optimization and risk management.
The main motivation for this analysis was the apparent contradiction
between results obtained on the basis of random matrix theory and the
fact that covariance matrices are so widely utilized for investment
or risk management purposes.
Using a simulation-based approach we have shown that for parameter
values typically encountered in practice the effect of noise on the
risk of the optimal portfolio may not necessarily be as large as one might
expect on the basis of the results of \cite{galluccio,bouchaud,stanley}.
The large discrepancy between ``predicted'' and ``realized''
risk obtained in \cite{bouchaud2,stanley2} can be explained by the
low values of $T/n$ used in these studies; for larger $T/n$ the
effect becomes much smaller. The analytic formulae derived in this paper
provide a lower bound on the effect of noise, or, conversely, an upper
bound on the size of the portfolios whose risk can be estimated with a
prescribed error. Finally, we have shown that for portfolios with weights
determined
independently from the covariance matrix data, the effect of noise
on the risk measurement process is quite small.

A very interesting topic for further research would be to analyze
the magnitude of these effects for portfolios constructed by
techniques really used in practice, for example by an asset manager
or hedge fund, since these methods often combine sophisticated
optimization schemes with more subjective expert assessments.


\section*{Acknowledgements}

We are extremely grateful to G.\ Papp and M.\ Nowak for pointing out the
way to analytically calculate $q_0$ and for valuable discussions.
We are also obliged to J.-P.\ Bouchaud and L.\ Laloux for sharing with us
some of their data and for discussions.
This work has been supported by the Hungarian National Science
Found OTKA, Grant No.\ T 034835.




\end{document}